# Reentrant metallic transition at a temperature above $T_c$ at the breakdown of cooperative Jahn-Teller orbital order in perovskite manganites

Dipten Bhattacharya* and H.S. Maiti
*Electroceramics Division, Central Glass and Ceramic Research Institute, Calcutta 700 032, India*

We report an interesting reentrant metallic resistivity pattern beyond a characteristic temperature $T^*$ which is higher than other such characteristic transition temperatures like $T_C$ (Curie point), $T_N$ (Neel Point), $T_{CO}$ (charge order onset point) or $T_{OO}$ (orbital order onset point) in a range of rare-earth perovskite manganites ($RE_{1-x}A_xMnO_3$; RE = La, Nd, Y; A = Sr, Ca; x = 0.0-0.5). Such a behavior is normally observed in doped manganites with doping level (x) higher than the critical doping level $x_c$ (= 0.17-0.22) required for metallic ground state to emerge and hence in a system where cooperative Jahn-Teller orbital order has already undergone a breakdown. However, the observation made in the $La_{1-x}Ca_xMnO_3$ (x = 0.0-0.3) series turns out to be an exception to this general trend.

The series of experimental and theoretical works on the orbital physics in the family of rare-earth perovskite manganites has revealed quite a few very interesting features: orbital wave (or orbiton)[1] in parent $LaMnO_3$; orbiton-phonon interaction;[2] orbital order in presence of tilt order;[3,4] complex orbital order in doped systems[5] where long range order no longer exists etc. Like in other strongly correlated systems, the orbital physics seems to be playing a crucial role in manganites as well, in governing the charge transport and magnetic properties (including colossal magnetoresistivity) or in coupling of charge, spin and orbital degrees of freedom.[6] While it has often been seen and well reported that orbital order governs the $T_C$, $T_{CO}$ or $T_N$ across the phase space diagram, we report, in this paper, its correlation with the onset of reentrant metallic resistivity pattern at a higher temperature. We have observed a reentrant metallic resistivity pattern beyond a characteristic temperature $T^*$ in a range of compounds within the entire family of hole-doped manganites. $T^*$ is found to be varying systematically with the doping level (x) as well as average A-site radius $\langle r_A \rangle$. We attempt to identify the x-$\langle r_A \rangle$ range of the compounds for which one observes the reentrant metallic resistivity pattern within a reasonable temperature limit (~1473 K). We note that the presence of long-range orbital order is, in general, detrimental to the occurrence of reentrant metallic resistivity pattern. The observation made in the $La_{1-x}Ca_xMnO_3$ (x = 0.0-0.3) series turns out to be the sole exception.

All the experiments have been carried out on phase pure bulk polycrystalline samples. They have been prepared by usual technique of compaction of powder and subsequent heat treatment at 1300-1400$^o$C for 5-10h under air or Ar atmosphere. The powder is prepared by a solution chemistry route called autoignition of citrate-nitrate gel and characterized by X-ray diffraction (XRD), scanning electron microscopy (SEM), energy dispersive X-ray spectra (EDX) etc.[7] We have also estimated, from wet chemical analysis,[8] the exact amount of $Mn^{4+}$ concentration in a sample which may be different from the degree of A-site doping (x) because of nonstoichiometry. The values of the $Mn^{4+}$ concentration for the low-doped systems are mentioned in the Table-I. We have noticed that with the increase in the doping level beyond x $\approx$ 0.2, the difference between the doping level and the actual amount of $Mn^{4+}$ concentration drops down.

The resistivity is measured for all the samples over a wide temperature range 77 – 1473 K. High pure Pt leads and conductive Pt paste are used for making the contacts in four-probe configuration for the resistivity measurement at high temperature. The samples, after wiring, are cured at 1473 K overnight. The thermal expansion has also been measured over a temperature range 300 – 1473 K using a dilatometer (ORTON, Model 1600). Signature of orbital order-disorder transition is conspicuous in thermal expansion vs temperature patterns as well. However, no such signature could be located in association with the reentrant metallic pattern.

In Fig. 1, we show the resistivity ($\rho$) vs. temperature (T) plots for a few selected samples over a temperature range 300 – 1473 K. Only the high temperature part is shown here for clarity. Following features can be noted from these plots: (i) in a few compositions (mainly parent and low-doped manganites), a distinct drop in resistivity takes place at around $T_{JT}$ (onset of orbital order due to cooperative Jahn-Teller effect); no further feature is present above $T_{JT}$; the $T_{JT}$ of undoped $LaMnO_{3+\delta}$ system appears to be somewhat lower because of nonstoichiometry ($Mn^{4+}$ concentration $\approx$ 5%) which results in a drop in $T_{JT}$; $T_{JT}$ also drops sharply with the increase in 'Sr' doping; however, $T_{JT}$ of the 'Ca'-doped samples, depicts very little doping dependence right up to a doping level x = 0.3 even though the change in entropy ($\Delta S_{JT}$) associated with the transition is suppressed with the increase in 'x';[4] (ii) in another set of doped manganites (with doping level $x \geq x_c$), metal-insulator transition takes place at $T_C$ as well as reentrant metallic state develops above $T^*$; no characteristic drop in resistivity at $T_{JT}$ is present; (iii) the data of $La_{1-x}Ca_xMnO_3$ (x = 0.0-0.3) (LCMO) series, however, reveal both a distinct abrupt drop in resistivity at $T_{JT}$ as well as a reentrant metallic pattern at $T^*$ ($T^* > T_{JT}$) for a doping level $x > x_c$ which appears to be an exception.

In Fig. 2, we show the thermal expansion ($\Delta l/l$; l = length of a sample) vs temperature plots for a few selected samples. Distinct kinks could be located in the patterns of low-doped samples around $T_{JT}$ which mark a sharp increase in thermal expansion coefficient '$\alpha$' in an orbital-disordered state. No such kink could be located at $T^*$.

For the purpose of our discussion, we categorize the whole set of samples depending on their doping level (x) and $<r_A>$ thus : (i) low x and low $<r_A>$ (x = 0.0-0.15; $<r_A>$ = 1.075-1.216 ) [category-I]; (ii) low x and high $<r_A>$ (x = 0.0-0.15; 1.216-1.235 ) [category-II]; (iii) moderate x and low $<r_A>$ (x = 0.15-0.3; $<r_A>$ = 1.0918-1.2103 ) [category-III]; LCMO series lies near the higher $<r_A>$ boundary of the entire range of category-III; (iv) moderate x and high $<r_A>$ (x = 0.15-0.3; $<r_A>$ = 1.235-1.245 ) [category-IV]; (v) high x and low $<r_A>$ (x = 0.3-0.5; $<r_A>$ = 1.1076-1.2048 ) [category-V]; once again, LCMO series lies near the higher $<r_A>$ boundary of the entire range of category-V; and (vi) high x and high $<r_A>$ (x = 0.3-0.5; $<r_A>$ = 1.245-1.263 ) [category-VI]. See Table II for the full list of compounds with their doping level (x) and $<r_A>$ values. In spite of little discrepancy between the doping level (x) and actual $Mn^{4+}$ concentration in a sample, we use, for convenience, the doping level (x) for qualitatively distinguishing the high temperature phase transitions across a wide spectrum of compositions.

A phase space diagram is drawn where variation of $T_{JT}$ and $T^*$ is shown as functions of average A-site radius $<r_A>$ and doping level (x) [Fig.3]. For clarity, we do not show other transition temperatures like $T_C$, $T_{CO}$, or $T_N$ here. It is interesting to note that while a finite $T_{JT}$ (but no $T^*$) is identifiable in the transport and thermal behaviors of the samples belonging mostly to the categories I and II, reentrant metallic resistivity pattern at $T^*$ (but no $T_{JT}$) could be observed for samples belonging primarily to the categories IV (when $x>x_c$) and VI. From the pattern of variation of $T^*$ with 'x' and '$<r_A>$', it can also be noted that $T^*$ is identifiable (within a range below ~1473 K) for those compounds of category-III (e.g., $Nd_{0.7}Sr_{0.3}MnO_3$, $Y_{0.7}Sr_{0.3}MnO_3$ etc.) for which finite $T_C$ yet no finite $T_{JT}$ could be located. $T^*$ drops towards $T_C$ for samples belonging to category-VI. And, in general, $T^*$ cannot be located in the resistivity patterns of the samples of category-V. The LCMO series of category-III, however, exhibits both $T_{JT}$ as well as $T^*$ (when $x>x_c$) and that of category-V exhibits $T^*$. The non-monotonic variation of $T^*$ with either x or $<r_A>$ underscores the fact that $T^*$ depends on both of them simultaneously.

From these observations, one can conclude that, in general, the presence of orbital order prevents the occurrence of reentrant metallic behavior. The orbital order can be controlled by simultaneously tuning the concentration of Jahn-Teller active $Mn^{3+}$ ions (1-x) and <$r_A$>. Breakdown of long-range orbital order is a necessary precondition for the observation of reentrant metallic resistivity pattern at $T^*$.

Unlike $T_{JT}$ or other such characteristic transition temperatures, $T^*$ does not mark a distinct phase transition. Rather it marks the completion of a crossover process set in at a much lower temperature or a relaxor-type transition process (often observed in spin glass or other relaxor systems[9]) taking place over a much wider temperature regime. The reason behind such a process is the lack of cooperative lattice distortion in a system having moderate doping and moderate to large <$r_A$>. In fact, doping leads to the formation of a frustrated matrix and, therefore, there are local transition temperatures instead of a global one. As a result of that, we observe no distinct kinks in the thermal expansion patterns (Fig. 2) of those compounds. Using an effective medium conductivity model $\sigma(T) = v_m(T)\sigma_m(T)+[1-v_m(T)]\sigma_{ins}$ where $\sigma_m(T)$ is the conductivity of the metallic phase, $\sigma_{ins}(T)$ is the conductivity of the insulating phase and $v_m(T)$ is the volume fraction of the metallic phase, one can show,[10] from the measured $\rho$ vs. T data, that $v_m(T)$ rises with temperature and $T^*$ actually marks the temperature at which $v_m(T)$ reaches 100%. It is to be noted here that due to the comparable conductivity values of both the insulating and metallic phases[11] in this high temperature regime, the overall charge conduction cannot be percolative. Therefore, only when the volume fraction of the metallic phase reaches 100% at the expense of the insulating phase, one observes such an onset of reentrant metallic resistivity pattern. The reason behind the local metallic transition and rise in $v_m(T)$ with temperature is not well known. This issue requires further work. It has been shown earlier by others[12] that small polaron hopping conduction dominates the charge conduction process in the absence of ferromagnetic exchange interaction across

Mn-O-Mn bonds above $T_C$. Our data suggest that, at least, for compounds belonging to moderate x high $<r_A>$ regime and beyond, such a conduction mechanism undergoes local transition progressively with temperature. Eventually, a global metallic pattern sets in at $T^*$. With the increase in doping and/or $<r_A>$, $T^*$ drops down and tends to $T_C$ (Fig. 3). In such a regime, instead of a paramagnetic insulating (PMI) phase at high temperature above $T_C$, one observes a paramagnetic metallic (PMM) phase. The cross-over from PMI to PMM phase with x is an important feature of the high temperature part of the phase diagram of hole-doped manganites.

In high x and small $<r_A>$ systems [category-V], the orbital order of $e_g$ level of Jahn-Teller ions is associated with $Mn^{3+}/Mn^{4+}$ charge order.[13] The charge/orbital order sets in at around 250-270 K ($T_{CO}$). Therefore, breakdown of long-range orbital order and simultaneous emergence of metallic ground state cannot be observed in this category of samples. And hence, no onset of re-entrant metallic pattern could be located in these systems within ~1473 K.

It is clear, then, that observation of $T^*$ within a reasonable temperature range is possible primarily within moderate x high $<r_A>$ to high x high $<r_A>$ range where metallic ground state and finite $T_C$ emerge at the breakdown of long-range Jahn-Teller orbital order.

The observation made in LCMO series is quite an exception. In the moderate doping range, both $T_{JT}$ as well as $T^*$ (for $x>x_c$) can be identified in the resistivity vs temperature plots while in the high doping range, $T^*$ could be located. It has earlier been reported[14] that doping in LCMO series leads to a phase segregation with the formation of both orthorhombic $O'$ (orbital-ordered) and O phases (orbital-disordered). Interestingly, $T_{JT}$ (measured from transport and thermal behaviors) depicts little drop due to such phase segregation. It seems possible as doping leads to a drop in $<r_A>$ in contrast to the case in 'Sr' doped systems.

In summary, the observation of a finite $T^*$ (a characteristic temperature beyond which reentrant metallic resistivity pattern can be observed) in a series of hole-doped manganites can be made at the breakdown of long-range Jahn-Teller orbital order and emergence of metallic ground state. With doping, first the $T_{JT}$ drops to virtually zero and finite $T_C$ emerges. With further doping, $T^*$ can be observed within a reasonable limit (~1473 K). $T^*$ eventually drops towards $T_C$ as doping level (x) and $<r_A>$ increase giving rise to a crossover from paramagnetic insulating (PMI) phase to paramagnetic metallic (PMM) phase. The $La_{1-x}Ca_xMnO_3$ (x = 0.0-0.3) series is an exception as very little drop in $T_{JT}$ could be observed with doping yet metallic ground state and reentrant metallic pattern develops beyond a doping level $x_c$.

The authors acknowledge the assistances rendered to them by Mr. S. Basu and Mr. A. Ghosh of the Institute.

*Present address : Department of Physics, Bar-Ilan University, Ramat Gan 52900, Israel; e-mail : dipten_bhattacharya_2000@yahoo.com

[1]E. Saitoh *et al.*, Nature **410**, 180 (2001).

[2]J. van den Brink, Phys. Rev. Lett. **87**, 217202 (2001).

[3]M.-v. Zimmermann *et al*., Phys. Rev. B **64**, 064441 (2001).

[4]D. Bhattacharya *et al.* (unpublished).

[5]J. van den Brink and D. Khomskii, Phys. Rev. B **63**, 140416 (R) (2001).

[6]See, for example, Y Tokura and N. Nagaosa, Science **288**, 462 (2000).

[7]A. Chakraborty, P.S. Devi, and H.S. Maiti, J. Mater. Res. **10**, 918 (1995).

[8]I.G.K. Andersen *et al*., J. Solid State Chem. **113**, 320 (1994).

[9]L. E. Cross, Ferroelectrics **76**, 241 (1987).

[10]D. Bhattacharya *et al*., J.Phys.:Condens. Matter **13**, L431 (2001).

[11]See, for example, A.K. Raychaudhuri, Adv. Phys. **44**, 21 (1995).

[12]M. Jaime *et al*., Phys. Rev. Lett. **78**, 951 (1997).

[13]See, for example, Y. Tomioka and Y.Tokura, in *Colossal Magnetoresistive Oxides*, edited by Y. Tokura (Gordon & Breach Science, Singapore, 2000) p. 281.

[14]See, for example, Y. Tokura, in *Colossal Magnetoresistive Oxides*, edited by Y. Tokura (Gordon & Breach Science, Singapore, 2000) p. 1; B.B. Van Aken *et al*., Phys. Rev. Lett. **90**, 066403 (2003).

Table-I. List of low-doped manganites and their relevant data

| Compositions | Doping Level (x) | $Mn^{4+}$ Concentration (%) | $T_{JT}$ (K) |
|---|---|---|---|
| $YMnO_{3+\delta}$ | 0 | 4 | 1273 |
| $NdMnO_{3+\delta}$ | 0 | 4 | 1050 |
| $LaMnO_{3+\delta}$ | 0 | 5 | 723 |
| $La_{0.9}Ca_{0.1}MnO_{3+\delta}$ | 0.10 | 15 | 733 |
| $La_{0.85}Ca_{0.15}MnO_{3+\delta}$ | 0.15 | 18 | 718 |
| $La_{0.975}Sr_{0.025}MnO_{3+\delta}$ | 0.025 | 10 | 673 |
| $La_{0.95}Sr_{0.05}MnO_{3+\delta}$ | 0.05 | 14 | 643 |
| $La_{0.9}Sr_{0.1}MnO_{3+\delta}$ | 0.10 | 17 | 458 |
| $La_{0.84}Sr_{0.16}MnO_{3+\delta}$ | 0.16 | 21 | ~0 |

Table-II. Full list of compounds used for constructing the phase space diagram and their relevant data

| Compositions | x | $\langle r_A \rangle$ (Å) | Category |
|---|---|---|---|
| $YMnO_3$ | 0 | 1.075 | I |
| $NdMnO_3$ | 0 | 1.163 | I |
| $LaMnO_3$ | 0 | 1.216 | I |
| $La_{0.9}Ca_{0.1}MnO_3$ | 0.10 | 1.2124 | I |
| $La_{0.85}Ca_{0.15}MnO_3$ | 0.15 | 1.2106 | I |
| $La_{0.975}Sr_{0.025}MnO_3$ | 0.025 | 1.2184 | II |
| $La_{0.95}Sr_{0.05}MnO_3$ | 0.05 | 1.2207 | II |
| $La_{0.9}Sr_{0.1}MnO_3$ | 0.10 | 1.2254 | II |
| $La_{0.84}Sr_{0.16}MnO_3$ | 0.16 | 1.231 | II |
| $La_{0.8}Ca_{0.2}MnO_3$ | 0.20 | 1.2088 | III |
| $La_{0.7}Ca_{0.3}MnO_3$ | 0.30 | 1.205 | III |
| $Nd_{0.8}Ca_{0.2}MnO_3$ | 0.20 | 1.1664 | III |
| $Nd_{0.7}Ca_{0.3}MnO_3$ | 0.30 | 1.1681 | III |
| $Y_{0.7}Ca_{0.3}MnO_3$ | 0.30 | 1.1065 | III |
| $Y_{0.7}Sr_{0.3}MnO_3$ | 0.30 | 1.1455 | III |
| $Nd_{0.8}Sr_{0.2}MnO_3$ | 0.20 | 1.1924 | III |
| $Nd_{0.7}Sr_{0.3}MnO_3$ | 0.30 | 1.2071 | III |
| $La_{0.8}Sr_{0.2}MnO_3$ | 0.20 | 1.235 | IV |
| $La_{0.75}Sr_{0.25}MnO_3$ | 0.25 | 1.2395 | IV |
| $La_{0.7}Sr_{0.3}MnO_3$ | 0.30 | 1.2442 | IV |
| $Y_{0.5}Ca_{0.5}MnO_3$ | 0.50 | 1.1275 | V |
| $Nd_{0.5}Ca_{0.5}MnO_3$ | 0.50 | 1.1715 | V |
| $La_{0.5}Ca_{0.5}MnO_3$ | 0.50 | 1.198 | V |

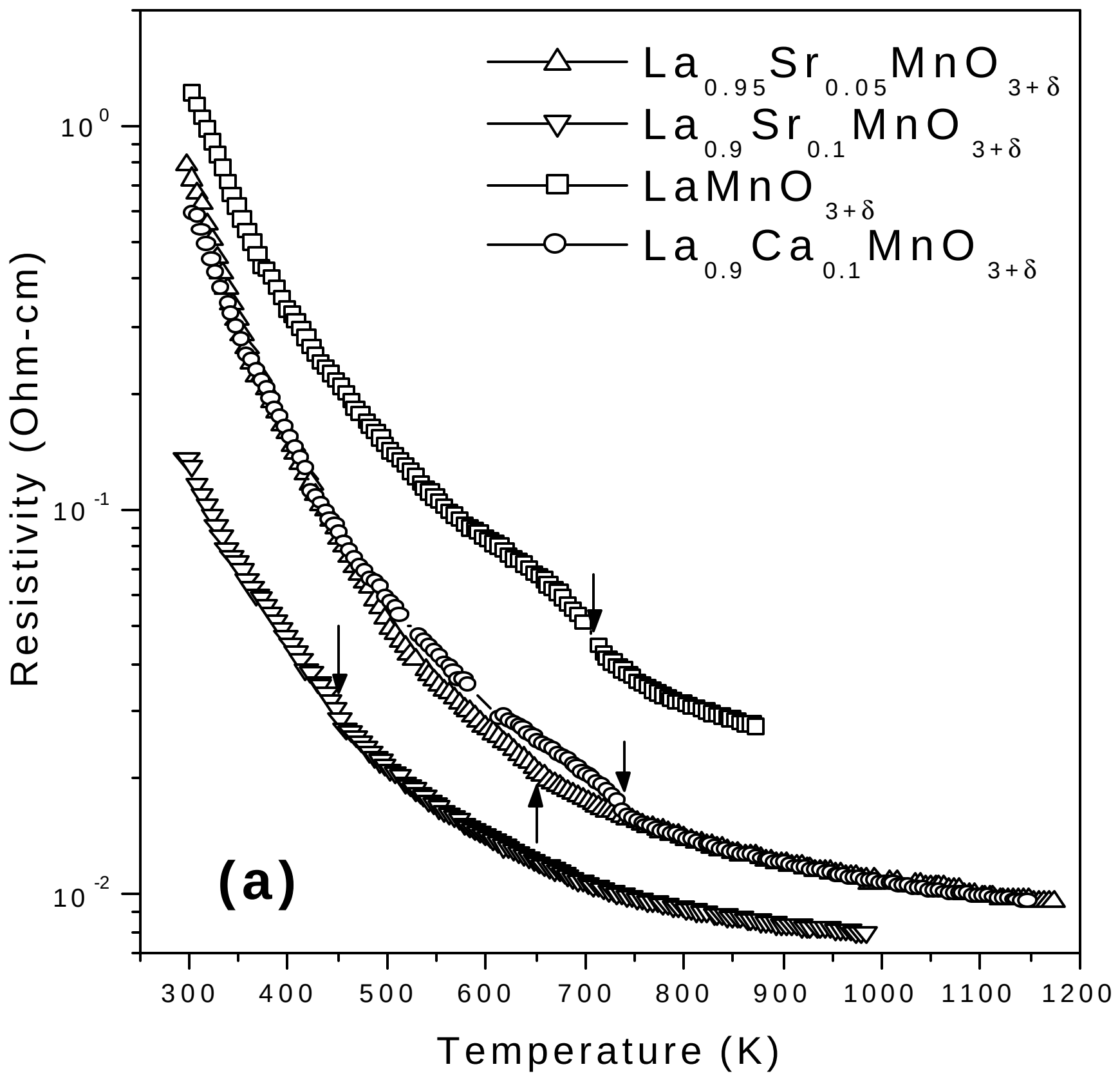

La0.95Sr0.05MnO3+δ
La0.9Sr0.1MnO3+δ
LaMnO3+δ
La0.9Ca0.1MnO3+δ
Resistivity (Ohm-cm)
10^0
10^-1
10^-2
(a)
300 400 500 600 700 800 900 1000 1100 1200
Temperature (K)

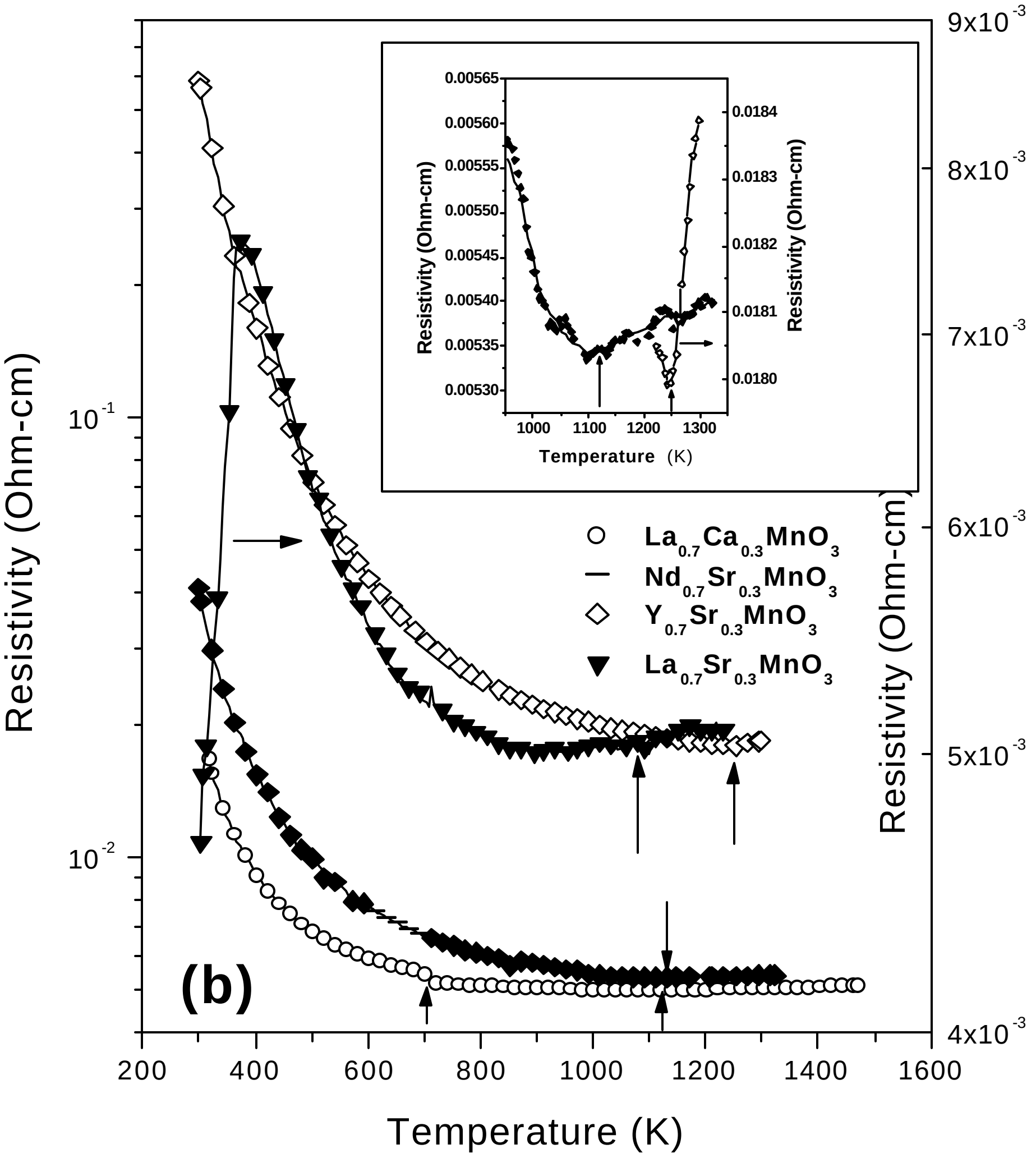

FIG. 1. (a) : Resistivity vs. temperature plots for low-doped systems - (i) undoped $LaMnO_{3+\delta}$; (ii) $La_{0.95}Sr_{0.05}MnO_{3+\delta}$; (iii) $La_{0.9}Sr_{0.1}MnO_{3+\delta}$; and (iv) $La_{0.9}Ca_{0.1}MnO_{3+\delta}$ - where distinct drop in resistivity around $T_{JT}$ could be observed. (b) : Resistivity vs. temperature plots for moderately doped systems - (i) $La_{0.7}Sr_{0.3}MnO_3$; (ii) $La_{0.7}Ca_{0.3}MnO_3$; (iii) $Nd_{0.7}Sr_{0.3}MnO_3$; and (iv) $Y_{0.7}Sr_{0.3}MnO_3$ - where re-entrant metallic patterns could be observed beyond $T^*$. Inset: the high temperature part of the resistivity plots for samples (iii) and (iv) is blown up to clearly show the metallic pattern

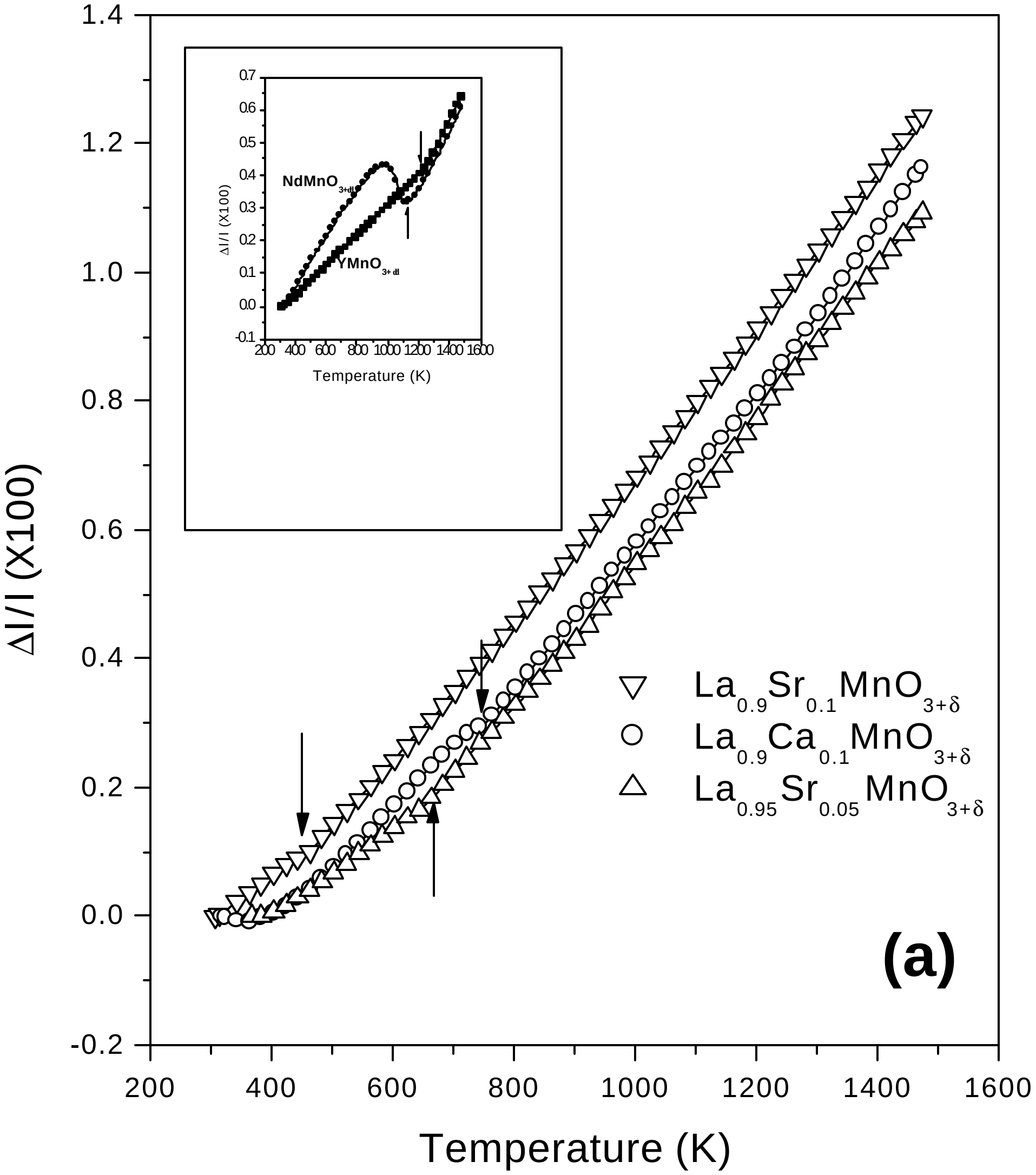

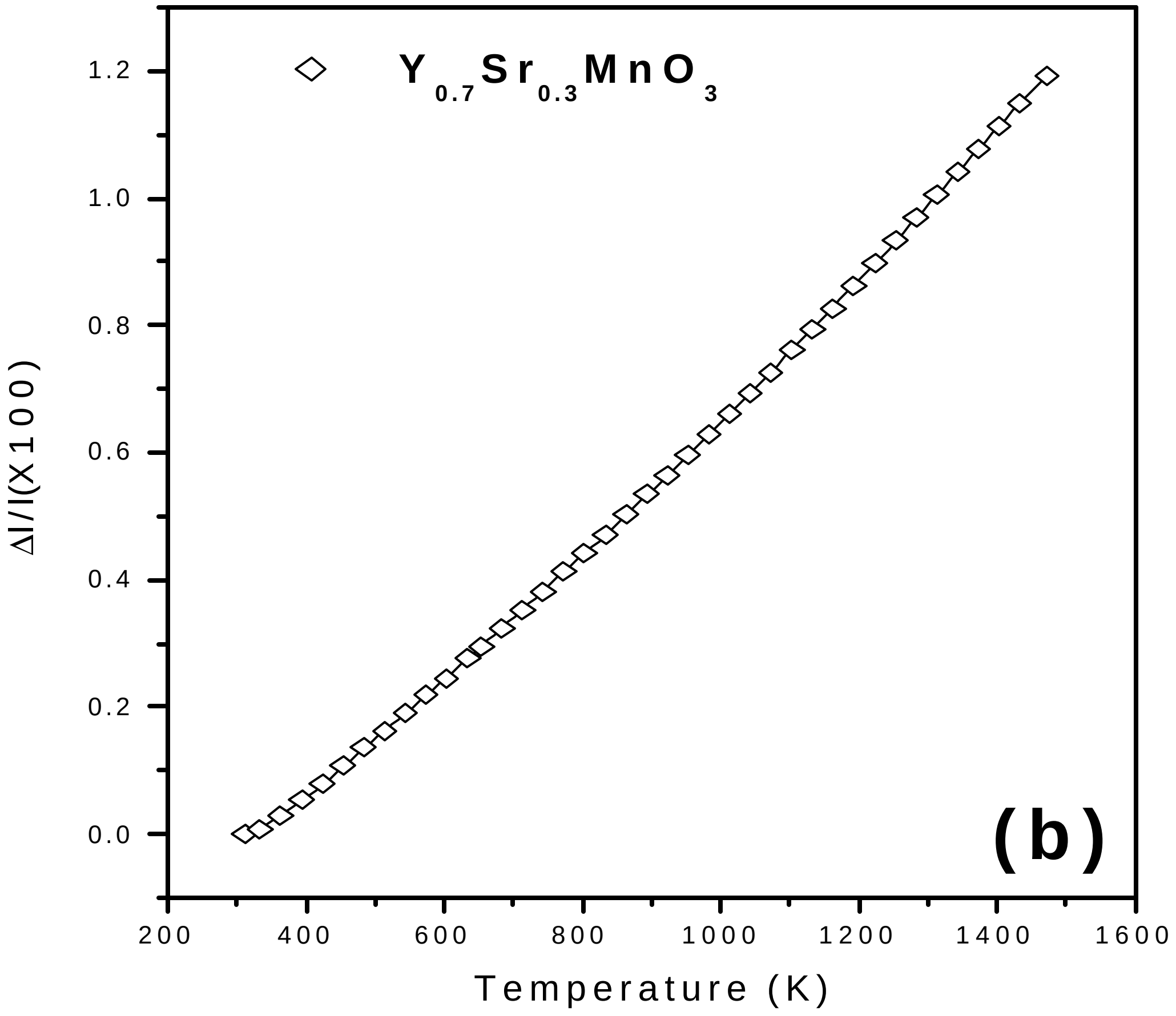

FIG. 2. (a) : The thermal expansion vs. temperature patterns for low-doped systems - (i) $La_{0.95}Sr_{0.05}MnO_{3+\delta}$; (ii) $La_{0.9}Sr_{0.1}MnO_{3+\delta}$; (iii) $La_{0.9}Ca_{0.1}MnO_{3+\delta}$ ; distinct kinks which mark change in thermal expansion coefficient ($\alpha$) could be located around $T_{JT}$. Inset: the thermal expansion patterns for undoped $NdMnO_3$ and $YMnO_3$ are shown. In these cases too, kinks could be observed at $T_{JT}$; (b) the thermal expansion vs. temperature pattern for a moderately-doped system $Y_{0.7}Sr_{0.3}MnO_3$; no distinct kink could be located at $T^*$.

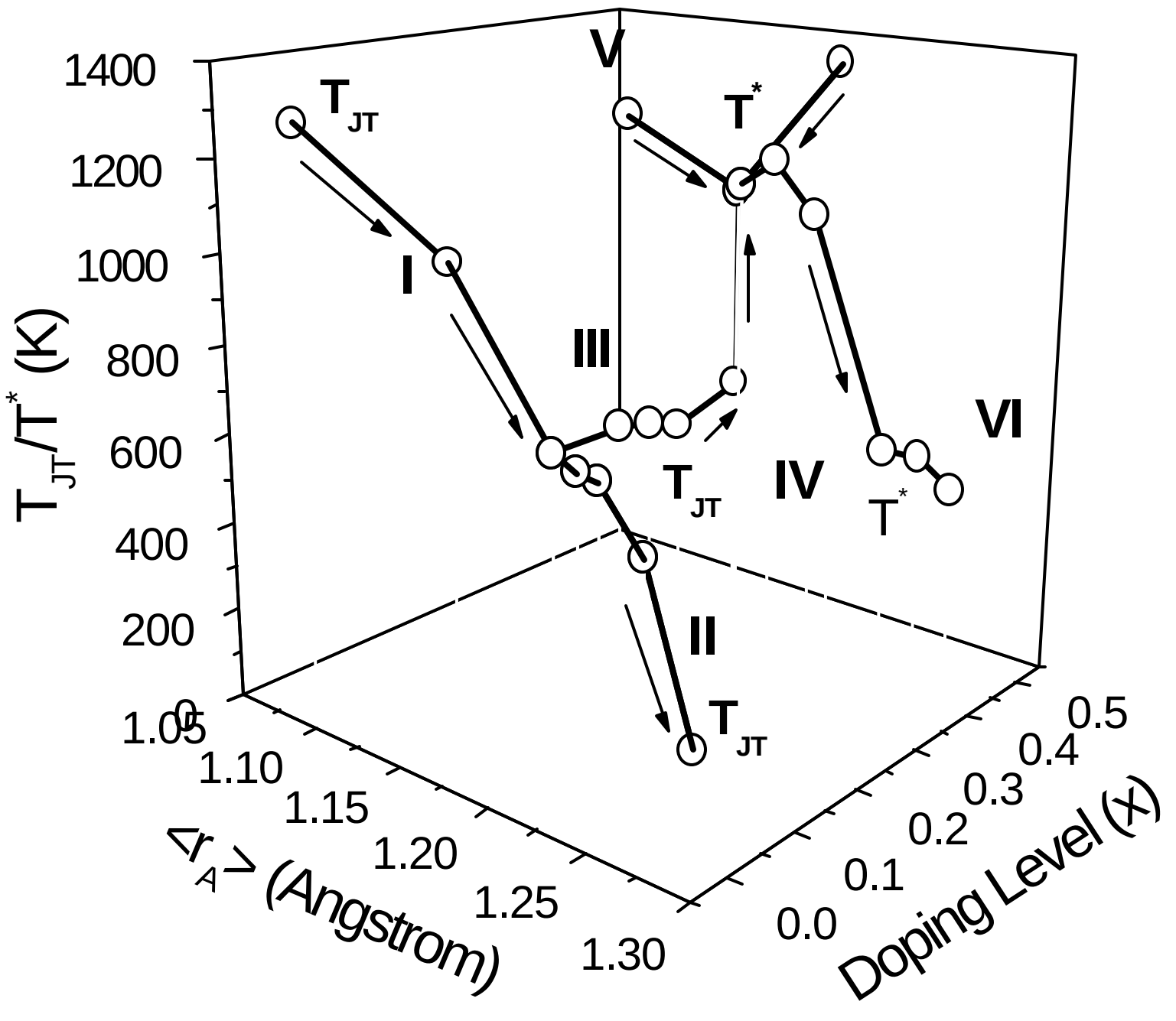

FIG. 3. The phase space diagram with variation in $T_{JT}$ and $T^*$ as a function of doping level (x) and average A-site radius <$r_A$>. $T_{JT}$ drops with doping in samples belonging to categories I and II; it remains more or less doping independent in the $La_{1-x}Ca_xMnO_3$ (x = 0.0-0.3) series belonging to category III. $T^*$, on the other hand, could be located in samples belonging primarily to categories IV, and VI; it drops with increase in doping and <$r_A$> values. Arrows for guiding the travel across the phase space diagram starting from the maximum $T_{JT}$.